%% file: SciPost_LaTeX_Template_Tau2018.tex
\documentclass[submission, Proceedings]{SciPost}
\usepackage{comment}
\input{authors}

\begin{document}

\begin{center}{\Large \textbf{Latest results of the OPERA experiment on nu-tau appearance in the CNGS neutrino beam}}\end{center}

\authorlist

\begin{center}
\today
\end{center}

\definecolor{palegray}{gray}{0.95}
\begin{center}
\colorbox{palegray}{
  \begin{tabular}{rr}
  \begin{minipage}{0.05\textwidth}
    \includegraphics[width=8mm]{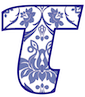}
  \end{minipage}
  &
  \begin{minipage}{0.82\textwidth}
    \begin{center}
    {\it Proceedings for the 15th International Workshop on Tau Lepton Physics,}\\
    {\it Amsterdam, The Netherlands, 24-28 September 2018} \\
    \href{https://scipost.org/SciPostPhysProc.1}{\small \sf scipost.org/SciPostPhysProc.Tau2018}\\
    \end{center}
  \end{minipage}
\end{tabular}
}
\end{center}


\section*{Abstract}
{\bf
OPERA is a long-baseline experiment designed to search for $\nu_{\mu}\to\nu_{\tau}$ oscillations in appearance mode. It was based at the INFN Gran Sasso laboratory (LNGS) and took data from 2008 to 2012 with the CNGS neutrino beam from CERN.
After the discovery of $\nu_\tau$ appearance in 2015, with $5.1\sigma$ significance, the criteria to select $\nu_\tau$ candidates have been extended and a multivariate approach has been used for events identification. In this way the statistical uncertainty in the measurement of the oscillation parameters and of $\nu_\tau$ properties has been improved. Results are reported.}

\vspace{10pt}
\noindent\rule{\textwidth}{1pt}
\tableofcontents\thispagestyle{fancy}
\noindent\rule{\textwidth}{1pt}
\vspace{10pt}

\section{Introduction}
\label{sec:intro}

Originated by the neutrino mass and the mixing between flavour and mass eigenstates, neutrino oscillations are now established thanks to intense experimental efforts. In 1998, the first evidence of neutrino oscillations was provided by the Super-Kamiokande experiment, showing the disappearance of atmospheric muon neutrinos~\cite{Kajita:1998nk2}. This result was consistent with the transition of $\nu_\mu$ to $\nu_\tau$ or to a new type of neutrino, still not known. At that time, moreover, the $\nu_\tau$ neutrino had not been observed yet. 

The OPERA experiment~\cite{OPERA_proposal} was designed to conclusively prove the existence of $\nu_\mu \to \nu_\tau$ oscillations. It was operated underground at the Gran Sasso INFN Laboratory (LNGS), 730 km away from the muon neutrino source at CERN, and collected data from 2008 to 2012.
The direct search for $\nu_\tau$ appearance was based on the detection of $\tau$ leptons produced in $\nu_\tau$ charged current interactions (CC).
The challenging detection of the short-lived $\tau$ lepton (c$\tau$ = 87~$\mu$m), out of almost twenty thousands $\nu_{\mu}$ interactions,
was achieved exploiting the nuclear emulsions sub-micrometric spatial resolution.

\section{The CNGS beam and the OPERA detector}

The OPERA detector 
was located at the underground Gran Sasso Laboratory (LNGS), 730~km away from the neutrino source, in the high energy CERN to LNGS beam (CNGS)~\cite{OPERA_4,opera4}. The average neutrino energy was $\sim17$~GeV, the $\bar{\nu}_{\mu}$ contamination was 2.1$\%$ in terms of interactions, the $\nu_{e}$ and $\bar{\nu}_{e}$ together were below 1$\%$, while the number of prompt $\nu_{\tau}$ was negligible.
The detector was a hybrid apparatus consisting of an emulsion/lead target complemented by electronic detectors. It was made up of two identical super-modules aligned along the CNGS beam direction, each made of a target section and a muon spectrometer. Each target section consisted of a multi-layer array of 31 target walls interleaved with pairs of planes of plastic scintillator strips. Target walls were made of Emulsion Cloud Chamber target units, called bricks, which were, in total, 150000.
Each brick
consists of 57 emulsion films, 300~$\mu m$ thick, interleaved with 56 lead plates, 1~mm thick.
The target total mass was 1.25~ktons.
The electronic detectors were used to identify the brick containing the neutrino interaction, for muon identification and its charge and momentum determination.

\section{Event selection and analysis}
Once a neutrino interaction was reconstructed in the electronic detectors, the bricks most probably containing the interaction vertex was identified by a dedicated offline algorithms and extracted from the walls. The nuclear emulsions were eventually developed and scanned to search for   
$\tau$ decays. 
The scanning was performed with automated optical microscopes installed in Laboratories in Europe and in Japan.
If a secondary vertex was found, a full kinematic analysis was performed combining the nuclear emulsion data with those from the electronic detectors. The momentum of charged particles in emulsions was determined by Multiple Coulomb Scattering~\cite{MCS:2011}.
For muons crossing the spectrometers, the momentum was measured with a resolution better than~22\% up to~30~GeV/c, and the charge sign determined~\cite{Agafonova:2011zz}.

The appearance of the $\tau$ lepton was identified by the detection of its characteristic decay topologies, either in one prong (electron, muon or hadron) or in three prongs. A first hint of a decay topology was the observation of an impact parameter larger than 10~$\mu$m, defined as the minimum distance between the track and the reconstructed vertex, excluding low momentum tracks. Kinematic selection criteria were then applied according to the decay channel.

\section{First phase of the OPERA experiment}

In the first phase of the OPERA experiment, very stringent kinematical selection criteria for $\nu_\tau$ candidate selection were applied, allowing a signal-to-background ratio of $\sim 10$.

Five $\nu_\tau$ candidates were observed: three in the $\tau\to 1h$ decay channel~\cite{OPERA:1tau, OPERA:4tau,OPERA:5tau}, one in the $\tau\to 3h$~\cite{OPERA:2tau} and one in the $\tau \to \mu$~\cite{OPERA:3tau} decay channel. In the sample analysed up to 2015, corresponding to 5408 neutrino interactions, $0.25\pm0.05$ background events were expected, coming mainly from events with an undetected primary muon, hadronic re-interactions 
and large angle muon scattering. 
The observation of five candidates results in $5.1\sigma$ significance for the exclusion of the background only hypothesis~\cite{OPERA:5tau}.

\section{Second phase of the OPERA experiment}

A new goal has been set. In order to estimate the oscillation parameters with reduced statistical uncertainty a new analysis procedure was implemented. 

Given the validation of the Monte Carlo simulation of $\nu_\tau$ events, based on different control data samples~\cite{Agafonova:2014khd, Ishida:2014qga, Longhin:2015dsa}, a new analysis strategy was developed, fully exploiting the features of expected $\nu_\tau$ events. A multivariate approach to improve signal to noise separation was applied to candidate events selected by means of moderately tight topological and kinematical cuts. 
The new selection was applied to the complete data sample, corresponding to 5603 $\nu$ interactions. 
Details about the new selection method are reported in~\cite{OPERA:final}. The total expected signal is $(6.8\pm 1.4)$ events, whereas the total background expectation is $(2.0\pm0.4)$ events.

\begin{figure}[h]
\centering
\includegraphics[scale=0.45]{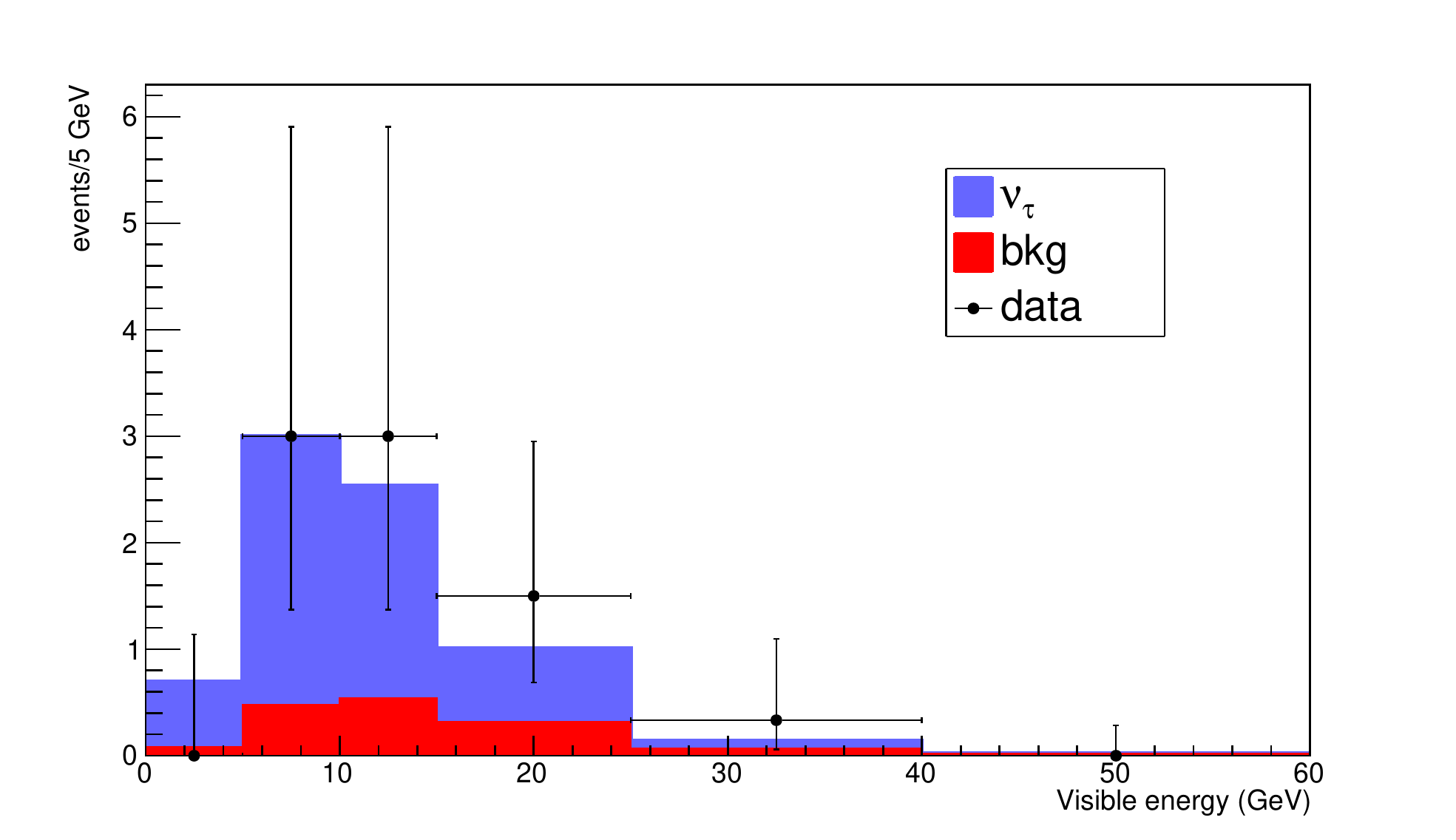}
\caption{Stacked plot of visible energy: data are compared with the expectation. Monte Carlo simulation is normalised to the expected number of events~\cite{OPERA:final}.}\label{fig:psum}
\end{figure}

Ten events ($N^{\textnormal{obs}}$) survived all the topological and kinematical cuts. The distribution of their visible energy, i.e.~the scalar sum of the momenta of charged particles and $\gamma$s, is shown in Fig.~\ref{fig:psum}, where it is compared to Monte Carlo simulation.

Different multivariate techniques have been considered and their performances for signal to background discrimination compared. The one with the best discrimination power was the Boosted Decision Tree (BDT).

\subsection{$\nu_\tau$ appearance statistical significance}

The statistical analysis used to re-evaluate the significance for the $\nu_\tau$ appearance is based on an extended likelihood constructed as the product of a probability density function given by the BDT response, a Poisson probability term which takes into account the number of observed events and the expected background in each decay channel, and a Gaussian term which accounts for systematics. The discovery significance of $\nu_\tau$ appearance is expressed in terms of a hypothesis test where the background only hypothesis plays the role of the null hypothesis and the signal-plus-background hypothesis is the alternative one. The null hypothesis was excluded with the improved significance of 6.1~$\sigma$~\cite{OPERA:final}.

\subsection{First measurement of $|\Delta m^2_{23}|$ in appearance mode and of $\nu_\tau$~CC cross-section on Lead}

The number of observed $\nu_\tau$ candidates after background subtraction is a function of the product of $\nu_\tau$ CC cross-section ($\sigma_{\nu_\tau}^{CC}$) and the oscillation parameter $\Delta m^2_{23}$. 

The squared mass difference $\Delta m^2_{23}$ was evaluated for the first time in appearance mode: assuming ${\sin^2 2\theta_{23}=1}$, $|\Delta m^2_{23}|$ is equal to ${(2.7^{+0.7}_{-0.6})\cdot 10^{-3}\textnormal{eV}^{2}}$. The result is consistent 
with the measurements performed in disappearance mode by other experiments and with the Particle Data Group best fit~\cite{PDG:2016}.

The $\nu_\tau$~CC cross-section on the OPERA lead target was also estimated: it is equal to ${(5.1^{+2.4}_{-2.0}) \cdot 10^{-36} \textnormal{cm}^2}$, assuming ${|\Delta m^2_{23}|=2.50\cdot 10^{-3}\ \textnormal{eV}^{2}}$. It is the first measurement of the $\nu_\tau$~CC cross-section with a negligible contamination from $\bar{\nu}_\tau$. 

\subsection{$\nu_\tau$ lepton number}

The OPERA experiment allowed to distinguish neutrinos from anti-neutrinos by the charge of the muon in $\tau$ muonic decays. This charge was determined as negative at 5.6~$\sigma$ level for the $\tau\to\mu$ candidate. Performing a dedicated BDT analysis which included also the background from $2\%$ $\bar\nu_\mu$ beam contamination, the first direct evidence for the leptonic number of $\tau$ neutrinos with a significance of $3.7 \sigma$ was obtained.

\section*{Conclusions}

OPERA claimed the discovery at $5.1\sigma$ of $\nu_\mu\to\nu_\tau$ appearance in the CNGS neutrino beam from the detection of five $\nu_\tau$ events, with a background of 0.25 events. A new analysis strategy was applied for the selection of additional $\nu_\tau$ candidates, in order to measure the oscillation parameters with reduced statistical error.


With the identification of five additional $\nu_\tau$ candidates, an overall sample of ten~$\nu_\tau$ candidates was collected, with~${2.0\pm0.4}$ expected background events. 
The discovery of $\nu_\mu\to\nu_\tau$ oscillations in appearance mode is confirmed with an improved significance of 6.1~$\sigma$.

Assuming ${\sin^2 2\theta_{23}=1}$, the first measurement of $|\Delta m^2_{23}|$ in appearance mode yields ${(2.7^{+0.7}_{-0.6})\cdot 10^{-3}\textnormal{eV}^{2}}$, while the measured $\nu_\tau$~CC cross-section on the lead OPERA target is ${(5.1^{+2.4}_{-2.0}) \cdot 10^{-36} \textnormal{cm}^2}$, assuming ${|\Delta m^2_{23}|=2.50\cdot 10^{-3}\ \textnormal{eV}^{2}}$. 

Furthermore, a dedicated BDT analysis in the $\tau\to\mu$ channel allows claiming for the first direct observation of the $\nu_\tau$ lepton number with a significance of 3.7~$\sigma$.

\section*{Acknowledgements}
We warmly thank  CERN for the successful operation of the CNGS facility and INFN for the continuous support given by hosting the experiment in its LNGS laboratory.

\paragraph{Funding information}
Funding is gratefully acknowledged from  national agencies and Institutions supporting us, namely: Fonds de la Recherche Scientifique-FNRS and Institut Interuniversitaire des Sciences Nucleaires for Belgium; MoSES for Croatia; CNRS and IN2P3 for France; BMBF for Germany; INFN for Italy; JSPS, MEXT, the QFPU-Global COE program of Nagoya University, and Promotion and Mutual Aid Corporation for Private Schools of Japan for Japan; SNF, the University of Bern and ETH Zurich for Switzerland; the Russian Foundation for Basic Research (Grant No. 12-02-12142 ofim), the Programs of the Presidium of the Russian Academy of Sciences (Neutrino Physics and Experimental and Theoretical Researches of Fundamental Interactions), and the Ministry of Education and Science of the Russian Federation for Russia, the Basic Science Research Program through the National Research Foundation of Korea (NRF) funded by the Ministry of Science and ICT (Grant No. NRF-2018R1A2B2007757) for Korea; and TUBITAK, the Scientific and Technological Research Council of Turkey for Turkey (Grant No. 108T324).



\bibliography{xbib.bib}

\nolinenumbers

\end{document}

%% file: authors.tex
\newcommand{\authorlist}{
\begin{center}
N. Agafonova\textsuperscript{1},
A. Alexandrov\textsuperscript{2},
A. Anokhina\textsuperscript{3},
S. Aoki\textsuperscript{4},
A. Ariga\textsuperscript{5},
T. Ariga\textsuperscript{5,6},
A. Bertolin\textsuperscript{7},
C. Bozza\textsuperscript{8},
R. Brugnera\textsuperscript{7,9},
A. Buonaura\textsuperscript{2,10}
% ...

\end{center}

\begin{center}
{\bf 1} INR - Institute for Nuclear Research of the Russian Academy of Sciences, RUS-117312 Moscow, Russia
\\
{\bf 2} INFN Sezione di Napoli, I-80126 Napoli, Italy
\\
{\bf 3} SINP MSU - Skobeltsyn Institute of Nuclear Physics, Lomonosov Moscow State University, RUS-119991 Moscow, Russia
\\
{\bf 4} Kobe University, J-657-8501 Kobe, Japan
\\
{\bf 5} Albert Einstein Center for Fundamental Physics, Laboratory for High Energy Physics (LHEP), University of Bern, CH-3012 Bern, Switzerland
\\
{\bf 6} Faculty of Arts and Science, Kyushu University, J-819-0395 Fukuoka, Japan
\\
{\bf 7} INFN Sezione di Padova, I-35131 Padova, Italy
\\
{\bf 8} Dipartimento di Fisica dell'Universit\`a di Salerno and ``Gruppo Collegato''  INFN, I-84084 Fisciano (Salerno), Italy
\\
{\bf 9} Dipartimento di Fisica e Astronomia dell'Universit\`a di Padova, I-35131 Padova, Italy
\\
{\bf 10} Dipartimento di Fisica dell'Universit\`a Federico II di Napoli, I-80126 Napoli, Italy
\\
{\bf 11} LPI - Lebedev Physical Institute of the Russian Academy of Sciences, RUS-119991 Moscow, Russia
\\
{\bf 12} JINR - Joint Institute for Nuclear Research, RUS-141980 Dubna, Russia
\\
{\bf 13} INFN - Laboratori Nazionali del Gran Sasso, I-67010 Assergi (L'Aquila), Italy
\\
{\bf 14} Dipartimento di Fisica dell'Universit\`a di Bari, I-70126 Bari, Italy
\\
{\bf 15} INFN Sezione di Bari, I-70126 Bari, Italy
\\
{\bf 16} LAPP, Universit\'e Savoie Mont Blanc, CNRS/IN2P3, F-74941 Annecy-le-Vieux, France
\\
{\bf 17} INFN Sezione di Bologna, I-40127 Bologna, Italy
\\
{\bf 18} IPHC, Universit\'e de Strasbourg, CNRS/IN2P3, F-67037 Strasbourg, France
\\
{\bf 19} Hamburg University, D-22761 Hamburg, Germany
\\
{\bf 20} Dipartimento di Fisica e Astronomia dell'Universit\`a di Bologna, I-40127 Bologna, Italy
\\
{\bf 21} Nagoya University, J-464-8602 Nagoya, Japan
\\
{\bf 22} GSSI - Gran Sasso Science Institute, I-40127 L'Aquila, Italy
\\
{\bf 23} Department of Physics, Technion, IL-32000 Haifa, Israel
\\
{\bf 24} METU - Middle East Technical University, TR-06800 Ankara, Turkey
\\
{\bf 25} INFN Sezione di Roma, I-00185 Roma, Italy
\\
{\bf 26} Ruder Bo\v{s}kovi\'c Institute, HR-10002 Zagreb, Croatia
\\
{\bf 27} Ankara University, TR-06560 Ankara, Turkey
\\
{\bf 28} Gyeongsang National University, 900 Gazwa-dong, Jinju 660-701, Korea
\\
{\bf 29} Center of Excellence for Advanced Materials and Sensing Devices, Ruder Bo\v{s}kovi\'c Institute, HR-10002 Zagreb, Croatia
\\
{\bf 30} Aichi University of Education, J-448-8542 Kariya (Aichi-Ken), Japan
\\
{\bf 31} Toho University, J-274-8510 Funabashi, Japan
\\
{\bf 32} Nihon University, J-275-8576 Narashino, Chiba, Japan
\\
{\bf 33} INFN - Laboratori Nazionali di Frascati dell'INFN, I-00044 Frascati (Roma), Italy
\\
{\bf 34} MEPhI - Moscow Engineering Physics Institute, RUS-115409 Moscow, Russia
\\
{\bf 35} Dipartimento di Fisica dell'Universit\`a di Milano-Bicocca, I-20126 Milano, Italy
\\
{\bf 36} HSE - National Research University Higher School of Economics, RUS-101000, Moscow, Russia
\\
{\bf 37} IIHE, Universit\'e Libre de Bruxelles, B-1050 Brussels, Belgium
\\
* giuliana.galati@na.infn.it
\end{center}
}